\documentclass[aps,prl,superscriptaddress,twocolumn]{revtex4} 
\usepackage{graphicx}
\usepackage{epstopdf}
\usepackage{color} 
\usepackage{soul} 

\begin{document}

\newcommand{\etal}{{\it et al.}\/}
\newcommand{\gtwid}{\mathrel{\raise.3ex\hbox{$>$\kern-.75em\lower1ex\hbox{$\sim$}}}}
\newcommand{\ltwid}{\mathrel{\raise.3ex\hbox{$<$\kern-.75em\lower1ex\hbox{$\sim$}}}}

\title{$\mathbf{d}$-wave and $\mathbf{s^\pm}$ Pairing Strengths in Ba$\mathbf{_2}$CuO$\mathbf{_{3+\mathbf{\delta}}}$}

\author{T.~A.~Maier}
\affiliation{Computational Sciences and Engineering Division, Oak Ridge National Laboratory, Oak Ridge, Tennessee 37831, USA}
\affiliation{Center for Nanophase Materials Sciences, Oak Ridge National Laboratory, Oak Ridge, TN 37831, USA}
\author{T.~Berlijn}
\affiliation{Computational Sciences and Engineering Division, Oak Ridge National Laboratory, Oak Ridge, Tennessee 37831, USA}
\affiliation{Center for Nanophase Materials Sciences, Oak Ridge National Laboratory, Oak Ridge, TN 37831, USA}
\author{D.~J.~Scalapino}
\affiliation{Department of Physics, University of California, Santa Barbara, CA 93106-9530, USA}

\date{\today}

\begin{abstract}
Using a first principles derived two-orbital model ($d_{x^2-y^2}$,
$d_{3z^2-r^2}$) and a random phase approximation treatment of the
spin-fluctuation pairing vertex, we calculate the $d$-wave and $s^\pm$-wave
pairing strengths for a model of the recently discovered
Ba$_2$CuO$_{3+\delta}$ superconductor. We find that there is significant
pairing strength in both of these channels. These results provide an
interesting perspective on the relationship between the cuprates and Fe-based
superconductors and the high-$T_c$ pairing mechanism.
\end{abstract}


\maketitle


In 2009, Geballe and Marezio \cite{ref:1} published a review of
superconductivity in Sr$_2$CuO$_{4-x}$ \cite{ref:2,ref:3} in which they noted
that, while it was isostructured to the familiar 214 La$_2$CuO$_4$ system, it
was extremely overdoped and had a superconducting transition temperature of
$T_c=95$~K, which was more than twice that of optimally doped
La$_{1.84}$Sr$_{0.16}$CuO$_4$. Since that time bulk superconductivity at
$T_c=84$~K in Cu$_{0.75}$Mo$_{0.25}$Sr$_2$YCu$_2$O$_{7.54}$ \cite{ref:4} and
exceeding 70~K in Ba$_2$CuO$_{3+\delta}$ \cite{ref:5} have been reported. All
of these materials have been synthesized at high temperatures and pressures in
the presence of a strong oxidizing agent. They are characterized as highly
overdoped with a reduced Cu apical O spacing compared to typical cuprate
superconductors. Here, using a two-orbital ($d_{x^2-y^2}$, $d_{3z^2-r^2}$)
tight-binding model derived from first principles for a single layer of
Ba$_2$CuO$_4$ and a multi-orbital random phase approximation (RPA) calculation
\cite{ref:6,ref:7} of the pairing vertex, we find that there is both
significant $A_{1g} (d_{x^2-y^2})$ and $B_{1g} (s^\pm)$ pairing strength in
this overdoped cuprate material with a compressed octahedron structure.

\begin{figure}[htbp]
 \includegraphics[width=1.0\columnwidth]{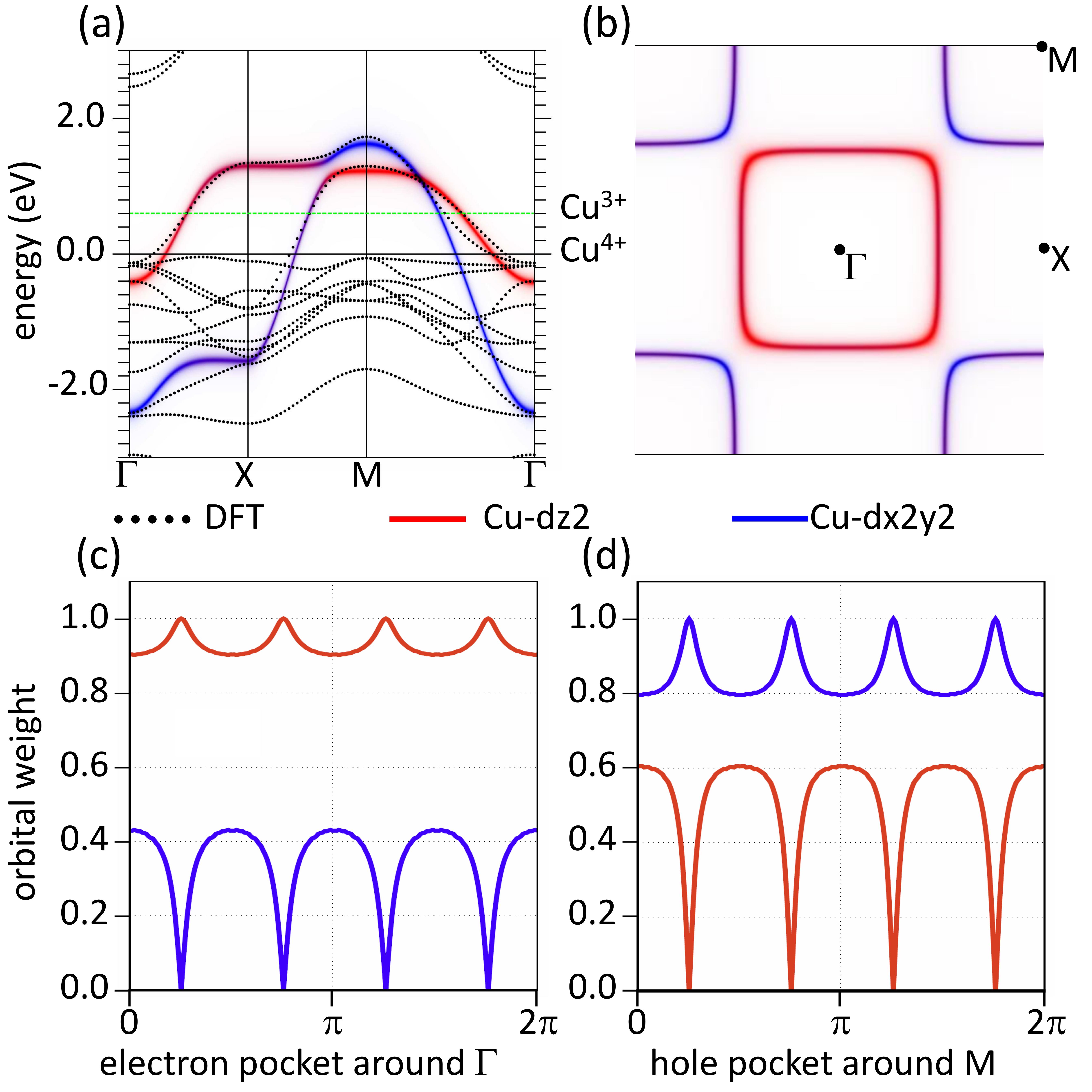} 
  \caption{\label{fig:1} (Color online) (a) The two-orbital Wannier function
and Density Functional Theory (DFT) bandstructures of Ba$_2$CuO$_4$. The green
line corresponds to a rigid band shift filling for Ba$_2$CuO$_{3.5}$. (b) The
Fermi surface for this filling with the orbital weight for $d_{x^2-y^2}$
(blue) and $d_{3z^2-r^2}$ (red). (c) and (d) $d_{x^2-y^2}$ and $d_{3z^2-r^2}$
orbital weights plotted versus the angle as $k$ varies around the Fermi
surfaces.}
\end{figure}

The density functional theory (DFT) bandstructure for a single layer of
Ba$_2$CuO$_4$ is shown in Fig.~\ref{fig:1}a. A Wannier transformation to a
2-orbital Cu-($d_{x^2-y^2}$, $d_{3z^2-r^2}$) model gives the two bands
illustrated in Fig.~\ref{fig:1}a. The Fermi surface assuming a rigid band
shift for Ba$_2$CuO$_{3.5}$ is shown in Fig.~\ref{fig:1}b. It consists of an
electron-like sheet with majority $d_{3z^2-r^2}$ orbital weight around the
zone center $\Gamma$ and a hole-like sheet with majority $d_{x^2-y^2}$ orbital
weight around the ($\pi,\pi$) point of the 2D Brillouin zone. The two orbital
Wannier function based Hamiltonian is given by
\begin{equation}
  H_0=\sum_{k\sigma}\sum_{\ell\ell'}\left(\xi_{\ell\ell'}\left(k\right)+
      \left(\varepsilon_\ell-\mu\right)\delta_{\ell\ell'}\right)
      d^\dagger_{\ell\sigma}(k)d^{\phantom\dagger}_{\ell'\sigma}(k)
\label{eq:1}
\end{equation} Here $\ell=1$ denotes the $d_{x^2-y^2}$ orbital and $\ell=2$
the $d_{3z^2-r^2}$ orbital. The tight-binding parameters and the technical
details of the first principles calculations are given in the Supplementary
Material. For the Fermi surfaces shown in Fig.~\ref{fig:1}b, the chemical
potential $\mu$ has been adjusted to give a doping $\delta=0.5$. The onsite
interaction part of the Hamiltonian has the usual form
\begin{eqnarray}
  H_1&=&U\sum_{i,\ell}n_{i\ell\uparrow}n_{i\ell\downarrow}+U'\sum_{i,\ell'<\ell}n_{i\ell}n_{i\ell'}\\
  \nonumber &+&J\sum_{i,\ell'<\ell\sigma,\sigma'}\sum
  d^\dagger_{i\ell\sigma}d^\dagger_
  {i\ell'\sigma'}d^{\phantom\dagger}_{i\ell\sigma'}d^{\phantom\dagger}_{i\ell'\sigma}\\
     \nonumber
     &+&J'\sum_{i,\ell'\ne\ell}d^\dagger_{i\ell}d^\dagger_{i\ell\downarrow}d^{\phantom\dagger}_
     {i\ell'\downarrow}d^{\phantom\dagger}_{i\ell'\uparrow}\,.
  \label{eq:2}
\end{eqnarray} A similar 2-orbital ($d_{x^2-y^2}$, $d_{3z^2-r^2}$) Hamiltonian
has been used by Jiang et al.~\cite{ref:8} to model a CuO$_2$ monolayer on
Bi2212 \cite{ref:9}. Here too the CuO$_2$ monolayer is heavily overdoped by
charge transfer at the interface and has a short Cu apical bond to the O in
the substrate. Treating this model using a Gutzwiller approximation, they
derived a spin-orbit superexchange pairing interaction of the Kugel-Kohmskii
form and find nodeless $A_{1g}$ $s^\pm$ pairing.

\begin{figure}[htbp]
 \includegraphics[width=1.0\columnwidth]{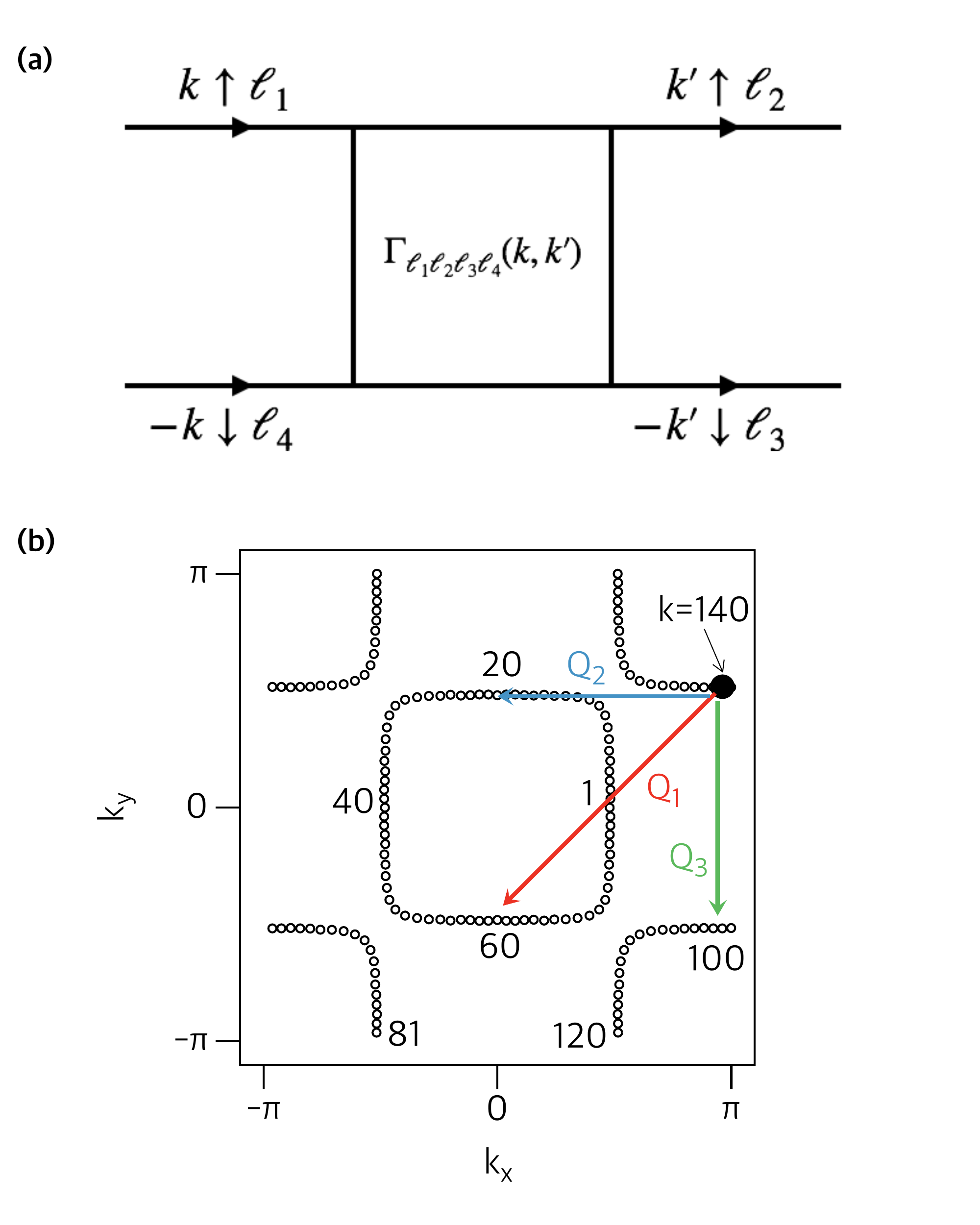}
  \caption{\label{fig:2} (Color online) (a) The scattering vertex
$\Gamma_{\ell_1\ell_2\ell_3\ell_4}(k,k')$ and (b) The point $k=140$ is fixed
at the bottom of the hole band and $k'$ varies along the Fermi surfaces. The
peaks in the vertices $\Gamma_{\ell_1\ell_2\ell_3\ell_4}(k,k')$ shown in
Fig.~\ref{fig:3} at $k'=60$ and the smaller ones at 20 and 100 arise from the
$k'-k$ scattering labeled $Q_1$, $Q_2$ and $Q_3$, respectively.}
\end{figure}

In the multi-orbital RPA theory the pairing vertex
$\Gamma_{\ell_1\ell_2\ell_3\ell_4}(k,k')$ for scattering a singlet pair
$(k\uparrow\ell_1,-k\downarrow\ell_4)$ in orbitals $\ell_1$ and $\ell_4$ to
$(k'\uparrow\ell_2,-k'\downarrow\ell_3)$ in orbitals $\ell_2$ and $\ell_3$ is
given by
\begin{eqnarray}
  \nonumber\Gamma_{\ell_1\ell_2\ell_3\ell_4}(k,k')&=&\left[\frac{3}{2}
  U^s\chi_S^{\rm RPA}(k-k') U^s \right.\\\nonumber &-& \frac{1}{2}\
  U^c\chi_O^{\rm RPA}(k-k')U^c \\ &+&\left.\frac{1}{2}\
  (U^s+U^c)\right]_{\ell_1\ell_2\ell_3\ell_4}
  \label{eq:3}
\end{eqnarray} Here $U^s$ and $U^c$ represent $4\times4$ matrices in orbital
space which depend on the interaction parameters and $\chi_S^{\rm RPA}$ and
$\chi_O^{\rm RPA}$ are orbital matrix RPA spin and orbital (charge)
susceptibilities given in the Supplementary Material.

In the following we will consider two sets of interaction parameters, the
first of which ($U=1$, $U'=0.5$, $J=J'=0.25$) satisfies rotational invariance
and the second ($U=U'=0.8$, $J=J'=0.4$) of which does not. These interaction
parameters like the tight-binding parameters given in the Supplemental
Material are given in eV. We will find for both sets that the $B_{1g}$
($d$-wave) and $A_{1g}$ ($s$-wave) pairing strengths are similar. For the
first parameter set the $B_{1g}$ strength is larger than that of the $A_{1g}$
channel but for the second parameter set the close balance between these two
is tipped in favor of the $A_{1g}$ channel.

\begin{figure}[htbp]
 \includegraphics[width=1.0\columnwidth]{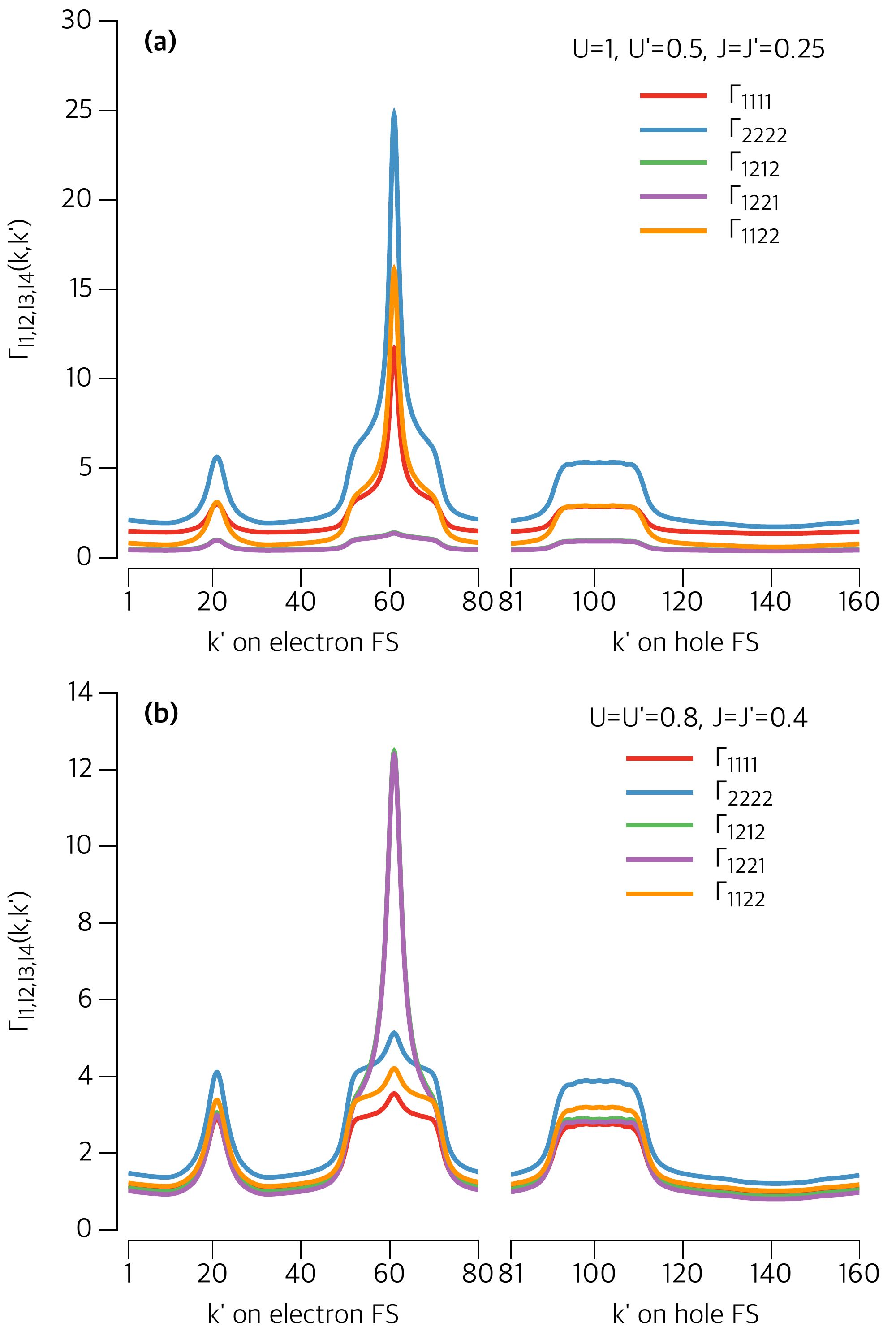} 
  \caption{\label{fig:3} (Color online) Selected orbital dependent vertices
  $\Gamma_{1111} (k,k')$, $\Gamma_{2222}(k,k')$, $\Gamma_{1212}(k,k')$ and
  $\Gamma_{1221} (k,k')$, $\Gamma_{1122}(k,k')$ for (a) $U=1.0, U'=0.5,
  J=J'=0.25$ and $T=0.06$, and (b) $U=U'=0.8, J=J'=0.4$ and $T=0.1$. In this
  case the $U'$ and $J'$ dependent scattering $\Gamma_{1221}(k,k')$ is
  largest. }
\end{figure}

The dominant orbital scattering vertices for the two different interaction
parameter sets are shown in Fig.~\ref{fig:3}a and \ref{fig:3}b.

\begin{figure}[thbp]
\includegraphics[width=1.1\columnwidth]{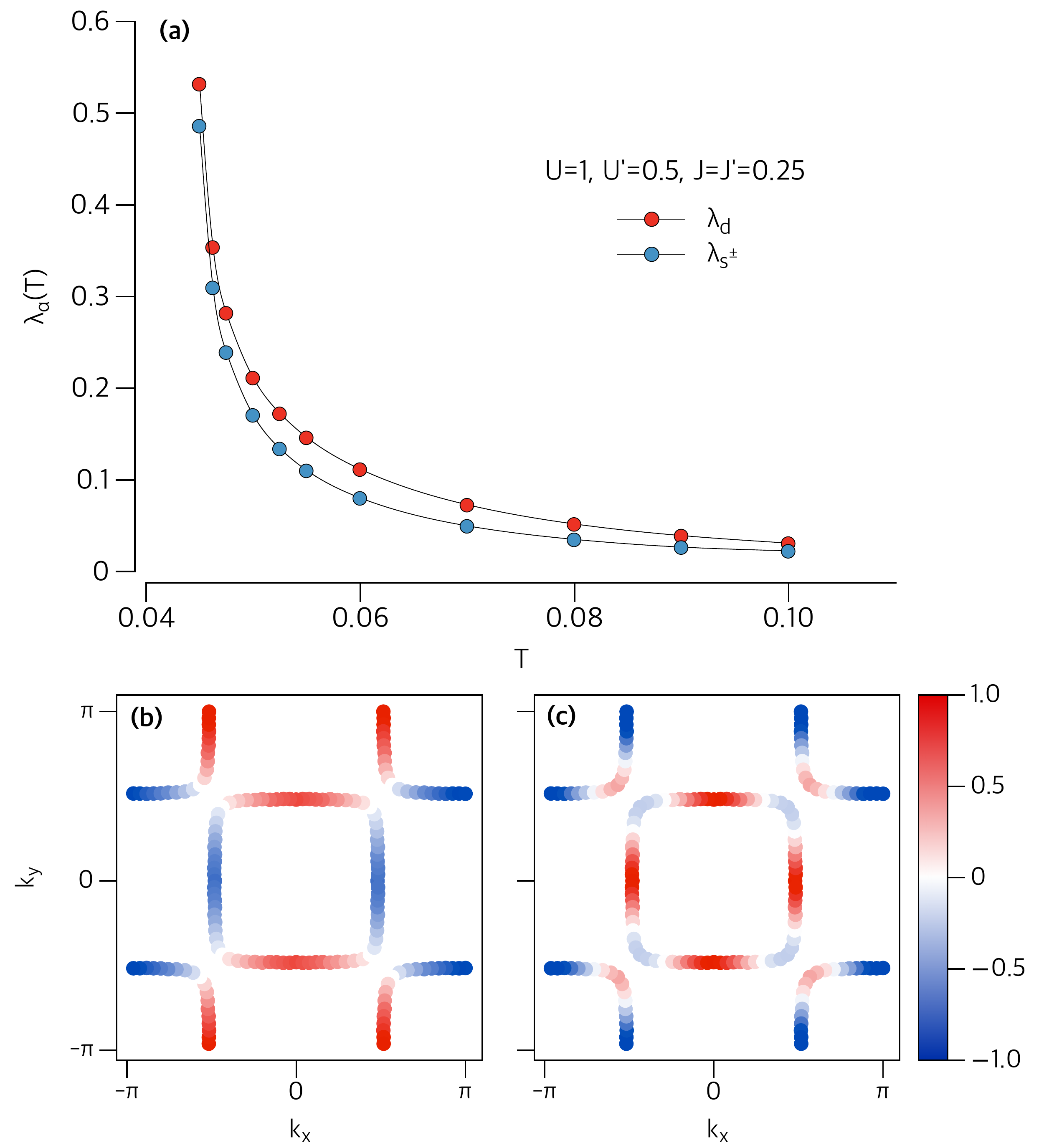}
  \caption{\label{fig:4} (Color online) (a) The pairing strength eigenvalues
  $\lambda_d(T)$ and $\lambda_{s^\pm}(T)$ versus temperature $T$ for $U=1.0,
  U'=0.5, J=J'=0.25$. The gap eigenfunctions $g_d(k)$ on the Fermi surface (b)
  for the $d$-wave at $T=0.06$ and (c) for the $s^\pm$-wave at $T=0.1$. }
\end{figure}
\begin{figure}[h]
 \includegraphics[width=1.1\columnwidth]{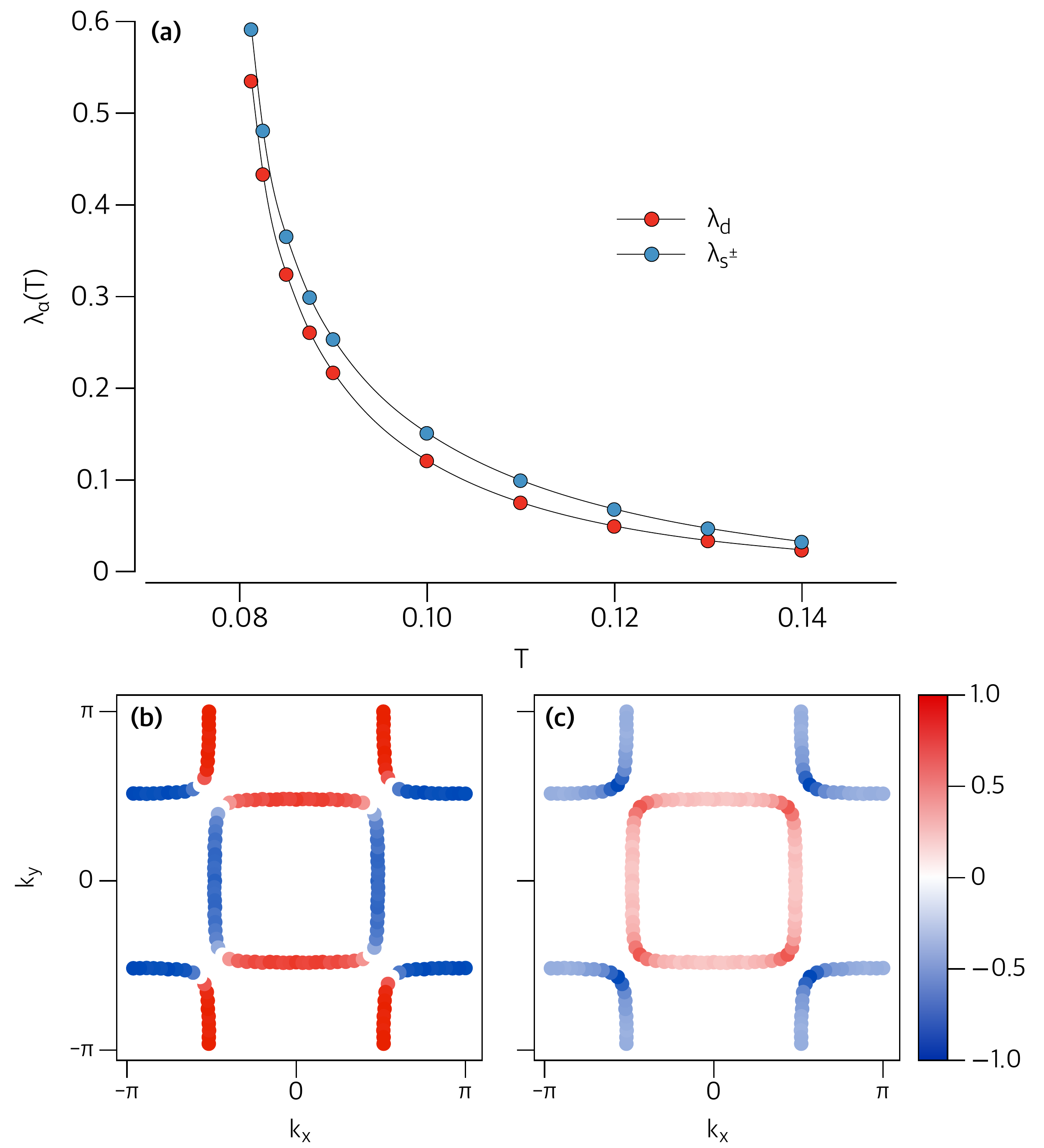}
  \caption{\label{fig:5} (Color online) (a) The pairing strength eigenvalues
  $\lambda_d(T)$ and $\lambda_{s^\pm}(T)$ versus temperature $T$ for
  $U=U'=0.8, J=J'=0.4$. The $d$-wave eigenfunction for $T=0.06$ and
  $s^\pm$-wave eigenfunction for $T=0.1$ on the Fermi surface are shown in (b)
  and (c) respectively.}
\end{figure}

Here $k$ is fixed at the bottom of the hole Fermi surface that surrounds the
$M$ point and $k'$ varies over the electron Fermi surface over points 1--80
and then over the hole Fermi surface points 81 to 160 as shown in
Fig.~\ref{fig:2}b. As shown in Fig.~\ref{fig:3}a for the rotationally
invariant parameter set, the dominant contribution to the pairing is
associated with the orbital vertices $\Gamma_{1111}$, $\Gamma_{2222}$,
$\Gamma_{1122}$, $\Gamma_{2211}$ in which the spin-up and spin-down electrons
remain in the same orbital states. In leading order these scattering processes
involve $U$ and $J$. In addition, there are scattering processes which involve
an orbital change such as $\Gamma_{1221}$ in which a pair in orbital
$d_{x^2-y^2}$ scatters to a pair in orbital $d_{3z^2-r^2}$ or $\Gamma_{1212}$
in which a pair $(k\uparrow 1,-k\downarrow 2)$ scatters to $(k\uparrow
2,-k\downarrow 1)$. These scattering processes involve $U'$ and $J'$ and
become significant for the second set of parameters shown in Fig.~\ref{fig:3}b
where the inter-orbital exchange $J'$ is larger.

Using these scattering vertices, the pairing strength is given by the
eigenvalue of
\begin{equation} -\sum_j\oint\frac{dk'_{\parallel}}{2\pi
  v_{F_j}(k'_{\parallel})}
  \Gamma_{ij}(k,k')g^\alpha_j(k')=\lambda_\alpha g^\alpha_i(k)
  \label{eq:4}
\end{equation} with \break
\begin{eqnarray}
  \Gamma_{ij}(k,k')&=&\sum_{\ell_1\ell_2\ell_3\ell_4}a^{\ell_1}_{\nu_i}(k)a^{\ell_4}_{\nu_i}(-k)
  \Gamma_{\ell_1\ell_2\ell_3\ell_4}(k,k')\nonumber\\
  &\times&a^{\ell_2*}_{\nu_j}(k')a^{\ell_3*}_{\nu_j}(-k')
  \label{eq:5}
\end{eqnarray} Here $j$ sums over the Fermi surfaces and
$v_{F_j}(k'_{\parallel})$ is the Fermi velocity $|\nabla_kE_{\nu_j}(k)|$. The
pairing strength eigenvalues for the $B_{1g}$ ($d$-wave) and $A_{1g}$
($s^\pm$-wave) along with the eigenfunctions $g_\alpha(k)$ which reflect the
expected structure of the superconducting gap are plotted in Fig.~\ref{fig:4}
and Fig.~\ref{fig:5} for the two parameters sets.

In both cases the pairing strengths for the $B_{1g}$ ($d$-wave) and the
$A_{1g}$ ($s^\pm$-wave) are quite close to each other. As seen in
Fig.~\ref{fig:2}b, the dominant pair scattering process for momentum transfer
($\pi,\pi$) involves the scattering shown as $Q_1$ which contributes
positively to the pairing strength in both the $d$-wave and $s^\pm$ channels.
Similarly the scattering $Q_2$ contributes positively to both, while the
scattering at $Q_3$ contributes negatively. The difference between the two
channels is tipped by a decrease in $U$ and an increase in $J'$.

To conclude, in a two-orbital ($3d_{x^2-y^2}$, $3d_{3z^2-r^2}$) RPA model of
Ba$_2$CuO$_{3+\delta}$ we find pairing strength in both the B$_{1g}$
($d$-wave) and A$_{1g}$ ($s$-wave) channels, with the $d$-wave channel favored
for rotationally invariant interaction parameters. The pairing strength is
large relative to a similar calculation for an optimally doped single-band
Hubbard model. This is in spite of the two orbitals each having significant
orbital weight at the Fermi energy. This appears counter to previous
calculations which have concluded that $T_c$ is optimized when the orbital
weight is concentrated in a single $3d_{x^2-y^2}$ orbital \cite{ref:P,ref:A}.
However, these calculations also found that the pairing strength depended upon
the shape of the Fermi surface and in the present case, the nearly square
shapes of both the electron and hole Fermi surfaces are responsible for the
enhanced pairing strength. We believe that this is what is responsible for the
second superconducting dome in the highly overdoped region of the cuprate
phase diagram proposed by Geballe and Marezio \cite{ref:1}.

The occurance of two pairing channels in this overdoped cuprate regime is
reminiscent of Ba$_{0.6}$K$_{0.4}$Fe$_2$As$_2$. In that case, below $T_c$ an
emergent B$_{1g}$ ($d$-wave) mode is observed in Raman scattering
\cite{ref:B}. This has been interpreted as arising from a Bardasis-Schrieffer
mode associated with a subdominant $d_{x^2-y^2}$ pairing channel in an $s^\pm$
superconductor. In the Ba$_2$CuO$_{3+\delta}$ case, one could have a $d$-wave
superconductor with a subdominant $s^\pm$ mode. The Raman observation of such
behavior would provide an interesting link between the cuprate and Fe-based
superconductors.

\section*{Acknowledgments}

The possibility of another region of superconductivity in the extremely
overdoped cuprates was proposed to one of us by Theodore H.~Geballe a number
of years ago. The DFT calculations, the derivation of the tight-binding model
(TB) and the analysis of the results (DJS) were supported by the Scientific
Discovery through Advanced Computing (SciDAC) program funded by U.S.
Department of Energy, Office of Science, Advanced Scientific Computing
Research and Basic Energy Sciences, Division of Materials Sciences and
Engineering. The RPA calculations (TAM) were supported by the U.S. Department
of Energy, Office of Basic Energy Sciences, Materials Sciences and Engineering
Division.


\end{document}


\newcommand{\etal}{{\it et al.}\/}
\newcommand{\gtwid}{\mathrel{\raise.3ex\hbox{$>$\kern-.75em\lower1ex\hbox{$\sim$}}}}
\newcommand{\ltwid}{\mathrel{\raise.3ex\hbox{$<$\kern-.75em\lower1ex\hbox{$\sim$}}}}

\title{$d$-wave and $s^\pm$ Pairing Strengths in Ba$_2$CuO$_{3+\delta}$ - Supplementary Material}

\author{T.~A.~Maier}
\affiliation{Computational Sciences and Engineering Division, Oak Ridge National Laboratory, Oak Ridge, Tennessee 37831, USA}
\affiliation{Center for Nanophase Materials Sciences, Oak Ridge National Laboratory, Oak Ridge, TN 37831, USA}
\author{T.~Berlijn}
\affiliation{Computational Sciences and Engineering Division, Oak Ridge National Laboratory, Oak Ridge, Tennessee 37831, USA}
\affiliation{Center for Nanophase Materials Sciences, Oak Ridge National Laboratory, Oak Ridge, TN 37831, USA}
\author{D.~J.~Scalapino}
\affiliation{Department of Physics, University of California, Santa Barbara, CA 93106-9530, USA}

\date{\today}


\date{\today}



\maketitle


\renewcommand{\theequation}{S\arabic{equation}}
\renewcommand{\thefigure}{S\arabic{figure}}


To derive the two-orbital Wannier function based Hamiltonian of Ba$_2$CuO$_
{3+\delta}$ we performed Density Functional Theory (DFT) calculations. To that
end we applied the WIEN2K\cite{wien2k} implementation of the full potential
linearized augmented plane wave method within the generalized gradient
approximation using the PBE exchange correlation functional.~\cite{pbe}. First
we performed a calculation of bulk Ba$_2$CuO$_{4}$ in which the space group
$I4/mmm$ and the lattice constants $a=7.5646$ Bohr and $c=24.4568$ Bohr were
taken from Ref.~\cite{CQJin} and in which the internal forces on the atoms were relaxed to less than 2 mRy/Bohr. The theoretically obtained apical O height of 3.65 Bohr compares well with the experimental value of 3.51 Bohr reported in Ref.~\cite{CQJin}. We used the theoretically obtained Ba and apical O heights together with the experimental in-plane lattice constant $a=7.5646$ Bohr to build a Ba$_2$CuO$_{4}$ monolayer with a vacuum of roughly 37 Bohr. A Wannier transformation of the monolayer calculation was performed with the \verb+wien2wannier+ and \verb+Wannier90+ software.~\cite{w2w,w90} Specifically a Wannier transformation was performed in which the Cu-$x^2-y^2$ and Cu-$z^2$ orbitals were projected on the bands within [-3.5,2] eV and in which the following \verb+Wannier90+ parameters were used \verb+num_iter=0+, \verb+dis_num_iter=1000+, \verb+dis_froz_min=0.5+ and \verb+dis_froz_min=2.5+. The elements of the resulting Wannier function based Hamiltonian were rounded up to the nearest meV, cut beyond 3rd nearest neighboring Cu atoms and are summarized in Tab.~\ref{tab:tab1}. Here, the
orbital matrices are labeled by $(\ell_1,\ell_2)$ with $\ell=1$ for the
$d_{x^2-y^2}$ orbital and $\ell=2$ for the $d_{3z^2-r^2}$ orbital. The k-point mesh was taken to be 5$\times$5$\times$5 for the bulk DFT calculation, 10$\times$10$\times$1 for the monolayer DFT calculation and 5$\times$5$\times$1 for the monolayer Wannier transformation. The basis set sizes of the DFT calculations were determined by \verb+RKmax=7+. 

\begin{table}[h!]
	\caption{Hopping parameters of the first principles Wannier function based Hamiltonian of Ba$_2$CuO$_{4}$ in eV. 
		\label{tab:tab1}}
	\begin{tabular}{c|c|c|c|c}
		\hline
		\hline
		 & on-site energy & 1$^{\rm st}$ ($t$) & 2$^{\rm nd}$ ($t^\prime$) & 3$^
		 {\rm rd}$
		 ($t^{\prime\prime}$)\\	\hline
		intra-orbital $d_{x^2-y^2}$ & $\varepsilon_1=0.624$ & $t_{11}=-0.495$
		& $t'_{11}=0.057$ & $t''_
				{11}=-0.099$ \\	\hline
		intra-orbital $d_{3z^2-r^2}$ & $\varepsilon_2=-0.185$ & $t_{22}=-0.205$ & $t'_{22}=-0.016$ &
		$t'_{22}=-0.039$ \\	\hline
		inter-orbital & 0 & $t_{12}=0.318$ & $t'_{12}=0$ & $t''_{12}=0.067$
		\\	\hline \hline
	\end{tabular}
\end{table}
With these parameters, the dispersions
$\xi_{\ell\ell'}(k)$ entering the Hamiltonian $H_0$ in Eq.~(1) are
\begin{eqnarray}\label{eq:H0}
\xi_{11/22}(k) &=& 2t_{11/22}(\cos k_x + \cos k_y)+4t'_{11/22}\cos k_x\cos
k_y\nonumber\\
&+&2t''_{11/22}(\cos 2k_x + \cos 2k_y)\nonumber\\
\xi_{12}(k)=\xi_{21}(k)&=&2t_{12}(\cos k_x - \cos k_y)+2t''_{12}(\cos 2k_x -
\cos
2k_y)
\end{eqnarray}

This Hamiltonian is then supplemented by the interaction terms given in Eq.~
(2) of the main text.  The
interaction matrices 
in the orbital basis $(11,22,12,21)$ are
\begin{eqnarray}
  U^S&=&\begin{pmatrix}U & J & 0 & 0 \\
	                     J & U & 0 & 0 \\
											 0 & 0 & U' & J' \\
											 0 & 0 & J' & U' \end{pmatrix}\\ \nonumber
  U^C&=&\begin{pmatrix}U & 2U'-J& 0 & 0 \\
	                     2U'-J & U & 0 & \\
											 0 & 0 & -U'+2J & J' \\
											 0 & 0 & J' & -U'+2J 
											 \end{pmatrix}\,.
  \label{eq:1}
\end{eqnarray}
The RPA spin $\chi^{\rm RPA}_S$ and charge $\chi^{\rm RPA}_O$ susceptibilities
are
\begin{eqnarray}
  \chi^{\rm RPA}_S(q)&=&[1-U^S\chi^0(q)]^{-1}\chi^0(q) \\ \nonumber
	\chi^{\rm RPA}_O(q)&=&[1-U^C\chi^0(q)]^{-1}\chi^0(q)
  \label{eq:2}
\end{eqnarray}
where the matrix elements of the bare susceptibility $\chi^0(q)$ are given by
\begin{equation}
  \chi^0_{\ell_1\ell_2\ell_3\ell_4}(q,\omega_m)=-\frac{T}{N}\sum_{k,\omega_n}
	G^0_{\ell_3\ell_1}(k+q,\omega_n+\omega_m)G^0_{\ell_2\ell_4}(k,\omega_n)
  \label{eq:3}
\end{equation}
with $G^0_{\ell\ell'}(k,\omega_n)$ the bare Green's function
\begin{equation}
  G^0_{\ell\ell'}(k,\omega_n)=\sum_\mu\frac{a^\ell_\mu(k)a^{\ell'*}_\mu(k)}
  {i\omega_n-\xi_\mu(k)}\,.
	\label{eq:4}
\end{equation}

\begin{figure}[htbp]
 \includegraphics[width=11.5cm]{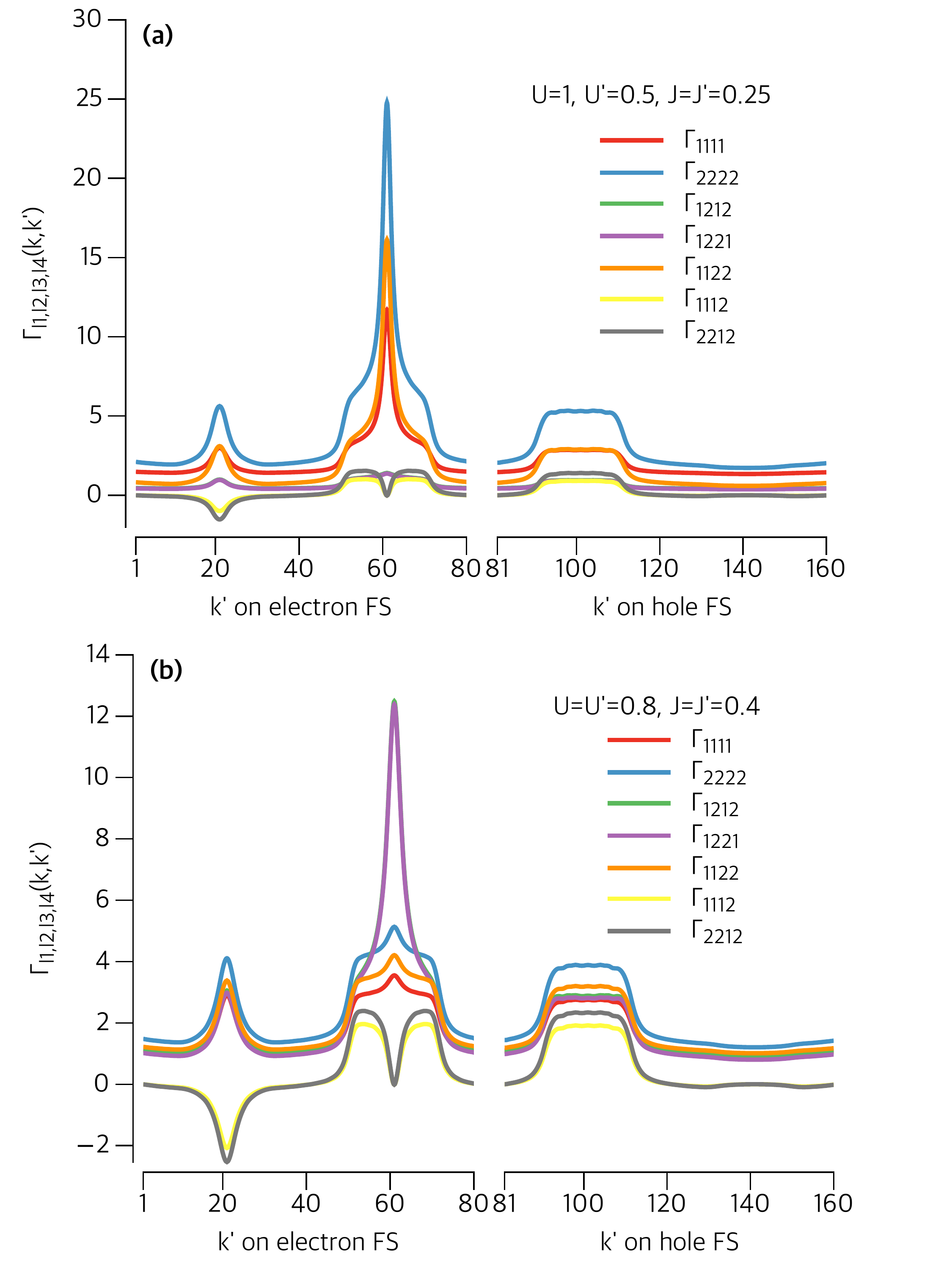}
  \caption{(Color online) The orbital vertices $\Gamma_
  {\ell_1\ell_2\ell_3\ell_4}(k,k')$ for $k$
	fixed as shown in Fig.~2(b) and $k'$ varying along the electron and hole Fermi
	surfaces. (a) $U=1.0$, $U'=0.5$, $J=J'=0.25$, $T=0.06$ and (b) $U=U'=0.8$,
	$J=J'=0.4$, $T=0.1$.\label{fig:1}}
\end{figure}

Typically the largest terms in the bare susceptibility are the diagonal $\chi^0_{1111}$,
$\chi^0_{2222}$, $\chi^0_{1212}$ and $\chi^0_{2121}$ terms and as a consequence
$\Gamma_{1111}$, $\Gamma_{2222}$, $\Gamma_{1122}$ and $\Gamma_{2211}$ depend on
the upper lefthand quadrant $U$, $J$ terms of the interaction while $\Gamma_
{1212}$,
$\Gamma_{1221}$, etc. depend upon $U'$ and $J'$. All of the orbital vertices
$\Gamma_{\ell_1\ell_2\ell_3\ell_4}(k,k')$ are shown in Figures~\ref{fig:1}a and
\ref{fig:1}b for the two sets of interaction parameters that we have studied.
Here one can see that the dominant vertices are those shown in Figs. 2
and 3.

\begin{figure}[htbp]
 \includegraphics[width=11.5cm]{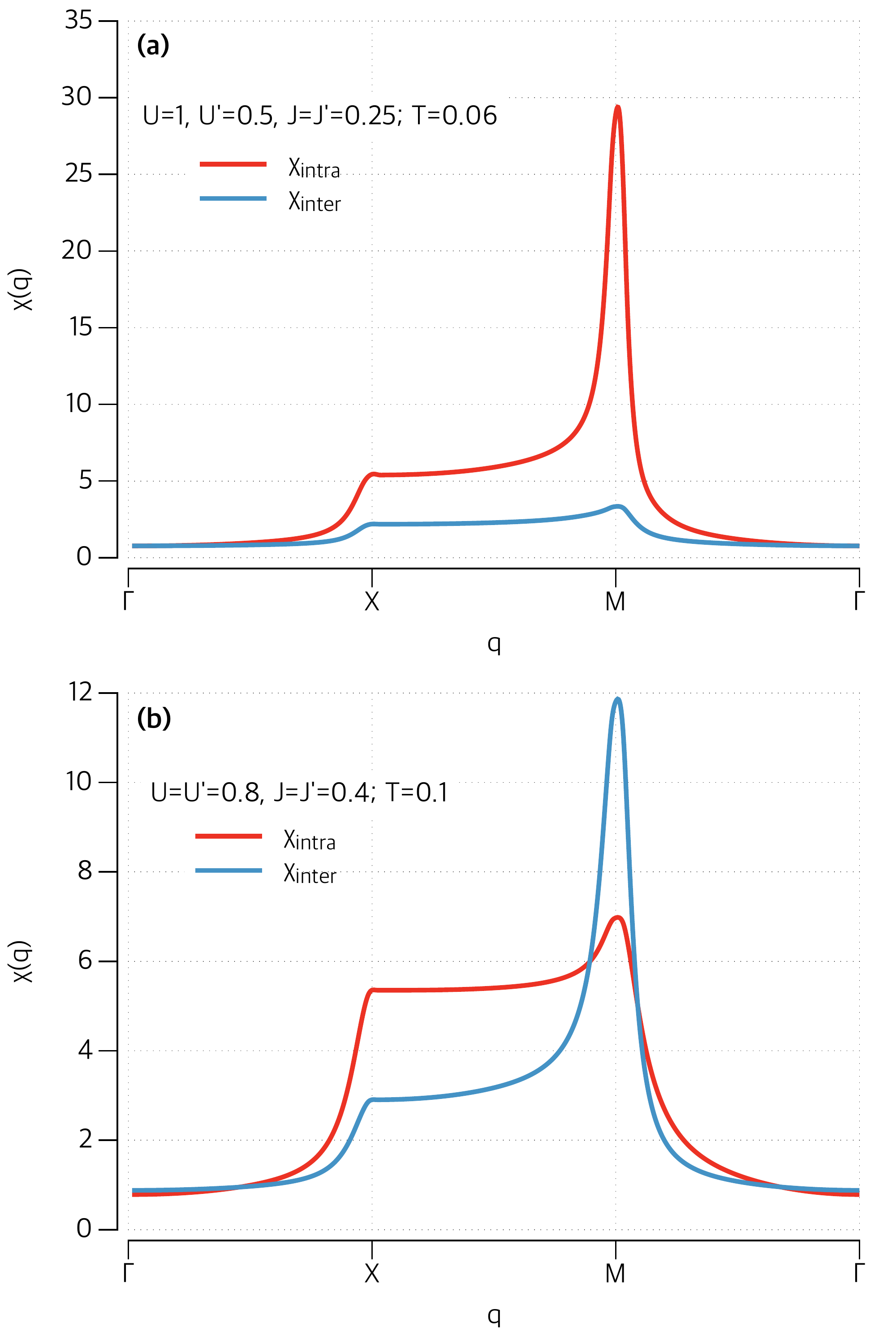}
  \caption{(Color online) The intra- and inter-spin susceptibility
	for (a) $U=1.0$, $U'=0.5$, $J=J'=0.25$ and $T=0.06$, and (b) $U=U'=0.8$,
	$J=J'=0.4$ and $T=0.1$.\label{fig:2}}
\end{figure}

As discussed in the text, the $\Gamma_{1212}$ and $\Gamma_{1221}$ vertices
involve the exchange of a fluctuation in which the orbital quantum number of the
up and down spins changes. In this case, the spin fluctuations which contribute
to the pairing are in the inter-orbital spin susceptibility channel. For example,
\begin{equation}
  \chi_{1212}(q,\omega_m)=\int^\beta_0d\tau e^{-i\omega_m\tau}
	\langle {\cal T}_\tau m^{\phantom\dagger}_{21}(q,\tau)m^\dagger_{21}(q,0)\rangle
	\label{eq:5}
\end{equation}
with
\begin{equation}
  m^\dagger_{21}(q)=\frac{1}{\sqrt N}\sum_kd^\dagger_{2\uparrow}(k+q)d^{\phantom\dagger}_{1\downarrow}(k)
	\label{eq:6}
\end{equation}
RPA results showing this inter-orbital spin susceptibility
\begin{equation}
  \chi_{\rm inter}(q)=\chi_{1212}(q,\omega_m=0)+\chi_{2121}(q,\omega_m=0)
  \label{eq:7}
\end{equation}
are plotted in Fig.~\ref{fig:2}a and Fig.~\ref{fig:2}b for the two parameter
sets we have studied.
RPA results for the usual intra-orbital spin susceptibility are also shown. For
the second set of interaction parameters the inter-orbital spin susceptibility
is clearly important. These fluctuations would not be seen in neutron
scattering.

\begin{figure}[htbp]
 \includegraphics[width=11.5cm]{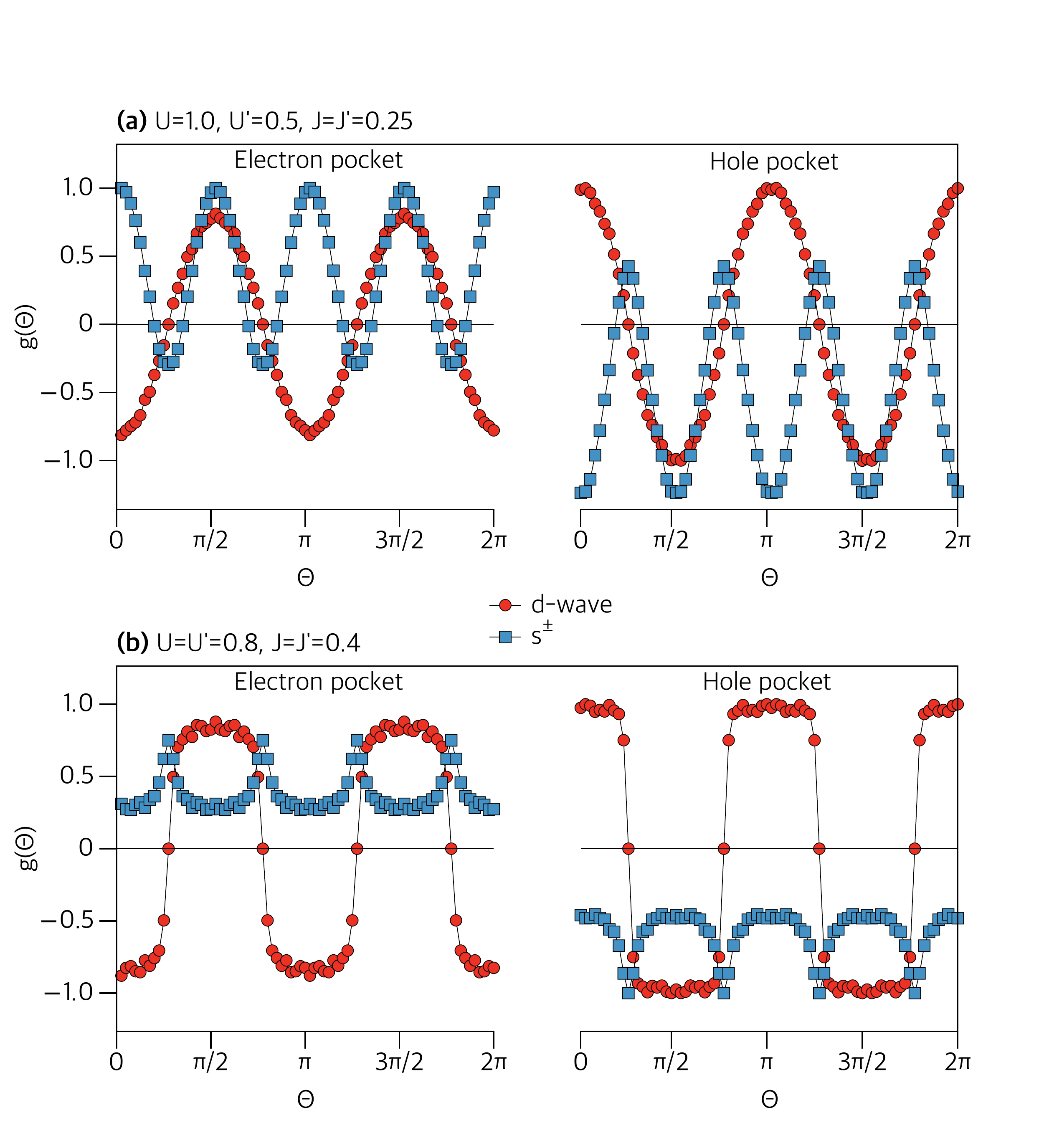}
  \caption{The pair eigenfunction $g_d(k)$ and $g_{s^\pm}(k)$ versus $k$ around the
	Fermi surfaces for (a) $U=1.0$, $U'=0.5$, $J=J'=0.25$ and $T=0.06$, and 
	(b) $U=U'=0.8$, $J=J'=0.4$ and $T=0.1$.\label{fig:s3}}
\end{figure}

Results for the pair eigenfunctions $g_d(k)$ and $g_{s^\pm}(k)$ versus $k$ are
shown in Fig.~\ref{fig:s3} for the two parameter sets discussed in our paper.
For the $U=1.0$ parameter set, the $s^\pm$ eigenfunction has 8 nodes, but for
the $U=0.8$ parameter set it is nodeless.



